\documentclass{article}
\usepackage{amsfonts}
\title{Nilpotent Fuzz}
\author{ Andrzej M. Frydryszak
\thanks{ Partially supported by the Polish KBN Grant \#
1PO3B01828.}\\
 Institute of Theoretical Physics,\\ University of Wroc{\l}aw,\\ pl.
Borna 9,\\ 50-204 Wroc{\l}aw, Poland\\ e-mail: amfry\@ ift.uni.wroc.pl \\[2ex]}
\begin{document}

\maketitle
\begin{abstract}
     We present a construction of the the formalism where fundamental variables are nilpotent, but in contrast to the supermathematics, commutative. 
     This gives another possibility to realize classically the Pauli exclusion principle. We sketch the relevant formalism and discuss simple model of nilpotent oscillator to illustrate the generalized nilpotent mechanics.
\end{abstract}

\section{Introduction}

The title of the present work origins from the expression used by Freed in \cite{fred}.
In that context fuzz denotes a kind of a 'cloud' surrounding a conventional geometrical object, like points, manifolds etc. In the case of supermanifold this fuzz is described by means of the Grassmannian parameters which are obviously nilpotent
\begin{equation}
\theta \theta'=-\theta' \theta \Rightarrow \theta^2=0
\end{equation}
However, he stresses that the nilpotency is more important property of such a 'fuzz'. Namely, when considering a ring of functions depending on ${\mathbb Z}_2$-variables %$C^{\infty}$
(let us denote it after Freed by  $C^{\infty}(W^1)$) he says: {\em "The fact that $C^{\infty}(W^1)$ is not a commutative algebra is a red herring; more important is the fact that it contains nilpotent elements"} (\cite{fred}, p. 20). The most natural step is to try a formalism where we admit commutative ring with nilpotent elements. \\
In the following we shall discuss an approach with nilpotent fuzz being abelian.
\begin{equation}
\eta \eta'=\eta' \eta \quad \mbox{and}\quad \eta^2=0
\end{equation}
As far as possible we will follow approach exploited in the supermathematics.
In the case of a commutative algebra we have to take into account various degrees of nilpotency, however we shall distinguish the first order nilpotents. \\
In the supersymmetric approach, classical anticommuting variables $\theta$ are objects used to realize 'classically' the Pauli exclusion principle i.e. $\theta^2=0$, but for them also relation to the statistic is given thanks to anticommutativity and anticommutators.\\
When considering the nilpotent but commuting fields we are to stay with the commutators as a natural operation, but relation to the exclusion principle is valid. As it was noted by Palumbo \cite{pal-1, pal-2, pal-3} this means that such fields should describe composite objects and formalism should allow to treat composites as independent entities without referring to components. It is interesting to observe that for the commuting nilpotent variables one can consider a version of the Berezin integral which for Gaussian function gives the permanent of a symmetric matrix. It is characteristic property. The term classical nilpotent mechanics is used here in the analogous sense as in pseudomechanics or classical supersymmetric mechanics \cite{berezin, casalbuoni}. In such theories we use classically nilpotent coordinates, but there is no, in the usual sense, classical observable that could be associated to them. It is a way of the pre-quantum description of systems which are the subject of the  Pauli principle. Such an approach turned out to be very useful in the description of spin, not only in the context of the Feynman path integral (cf. brief review \cite{af-rev}).\\
It turns out that nilpotent commuting variables are of use in LCFT (Logarithmic Conformal Field Theory), where they parametrize a nilpotent shift in the conformal weights \cite{lcft}. It allows to write single condition for two fields $\phi$, $\psi$ forming the Jordan cell i.e. for $\Phi(z, \eta)=\phi(z)+\eta\psi(z)$
\begin{equation}
\Phi(\lambda z, \eta)=\lambda^{-(\Delta+\eta)}\Phi(z, \eta)
\end{equation}
For components fields it reads as \cite{lcft}
\begin{equation}
\phi(\lambda z)= \lambda^{-\Delta}\phi(z), \quad\quad
\psi(\lambda z)=\lambda^{-\Delta}(\psi(z)-\phi(z)\ln \lambda)
\end{equation}
It appears that when using such variables and symmetry properties it is possible to derive many properties of LCFT e.g. two and three point correlation functions, Kac determinant \cite{lcft}.\\
Numbers with nilpotent part are known in mathematics for a long time. Generalized dual numbers were used by N. A. Gromov \cite{grom-0, grom-1, grom-2} in a series of papers devoted to the contractions of groups, quantum groups and description of spaces with degenerate metrics and relevant field theories. Earlier P. I. Pimenov \cite{pim} has given classification of spaces of the constant curvature
using nilpotent commuting numbers. Some other instances of the presence of such numbers are briefly described in the Ref. \cite{amf-dn}.\\
In the next section we will recall the definition of nilpotent variables and present the elements of differential and integral calculus. Its essential property is the lack of conventional Leibniz rule. We shall answer the question of the $\dot{\eta}^=0$. Then the formalism of the nilpotent mechanics in the Lagrangian and Hamiltonian form will be sketched.

\section{Nilpotent commuting variables}
It might seem naively that formalism with variables which are commuting and nilpotent should be easier to handle then $\mathbb{Z}_2$-graded one. However it turns out that it is not so. Demand of the nilpotency  for commuting variables is an additional strong condition, while for odd variables it is merely obvious conclusion from the $\mathbb{Z}_2$-graded commutativity. In the following we will use the definition of nilpotent variables formulated in Ref. \cite{amf-n2} (another definition of nilpotent numbers can be found in Ref. \cite{grom-0}).\\
Let us consider simple example of one variable case.
\begin{center}
\begin{tabular}{ccc}
Anticommutative&& Commutative\\
\hline\\
$G[\theta]$&\rule{3cm}{0cm}&$F[\eta]$\\
$g(\theta)=g_0+g_1\theta$&&$f(\eta)=f_0+f_1\eta$\\
$\theta^2=0$, $|\theta|=1$&&$\eta^2=0$, $|\eta|=0$\\
$\partial_{\theta}\theta=1$&&$\partial_{\eta}\eta=1$\\
$\partial_{\theta}\theta^2=\partial_{\theta}\theta\cdot\theta-\theta\cdot\partial_{\theta}\theta$
&&
$\partial_{\eta}\eta^2=\partial_{\eta}\eta\cdot\eta+\eta\cdot\partial_{\eta}\eta-2\eta$\\[3mm]
(Grassmann)&&(Paragrassmann)
\end{tabular}
\end{center}
Here on the level of functions we do not have any difference between both cases, only when we want to introduce a derivative the commutative and $\mathbb{Z}_2$-graded case are different. It turns out that for commuting variables, to be consistent with nilpotency, we have to modify the Leibniz rule \cite{amf-n2}.
\begin{equation}\label{mleib}
\partial(f\cdot h)= \partial f\cdot h + f \cdot \partial h -
2\eta\partial f\partial h
\end{equation}
Modification the Leibniz rule of this kind was considered in \cite{fik}.
This modification makes the new theory nontrivial. \\
To discuss mechanical systems described by the first order nilpotent coordinates $\eta$ we have to know properties of the time derivative for these variables. To this end let us consider a mapping $\eta: \mathbb{R}\mapsto \mathcal{D}$, where $\mathcal{D}\subset \mathcal{N}$ is the subset of the first order nilpotents, $\mathcal{D}=\{\eta\in\mathcal{N}\;|\; \eta^2=0\}$ and $\mathcal{N}$ is
the set of nilpotent numbers $\mathcal{N}$ introduced in \cite{amf-n2}. We shall call the $\eta(t)$  the first order nilpotent curve (the FON curve) of class $C^m$ if the real coefficient functions in the expansion
\begin{equation}\label{xiexpan}
\eta(t)=\sum_{k, I_k}\nu_{I_k}(t)\xi^{I_k}
\end{equation}
are of the class $m$ i.e. $\nu_{I_k}(t)\in C^m(\mathbb{R})$. We keep the notation from Ref.\cite{amf-n2} and the $\{\xi^{I_k}\}_{k,I_k}$ form the set of generators of $\mathcal{N}$. The time derivative of $\eta(t)$ is the function
\begin{equation}\label{velo}
\frac{d}{dt}\eta(t)=\sum_{k, I_k}\frac{d}{dt}\nu_{I_k}(t)\xi^{I_k},
\end{equation}
where the derivative on the right-hand side is the usual derivative of a real function.
For the FON curve we have from definition, that $\eta^2(t)=0$ for every $t$.
%and $\nu_{I_k}(t)\in C^m$.
Let us note  that the
segment joining two first order nilpotent points $\eta_1$, $\eta_2$
\begin{equation}
\eta(t)=(1-t)\eta_1+t\eta_2, \quad t\in[0,1].
\end{equation}
in general is not a FNC. Obviously $\eta^2(t)=2t(1-t)\eta_1\eta_2$ and it vanishes only for algebraically dependent $\eta_1$, $\eta_2$ i.e. when $\eta_1\eta_2=0$. From this we see that $\mathcal{D}$ is not a convex set.
For the $\eta(t)$ to be a FNC we have to fulfill condition
\begin{equation}
\eta^2(t)=\sum_{l,J_l}\sum_{k, I_k}\nu_{I_k}(t)\nu_{J_l}(t)\xi^{I_k}\xi^{J_l}=0
\end{equation}
This means that
 \begin{itemize}
 \item[(a)] $I_k\cap J_l\neq\emptyset$ for all $k$, $l$ and $I_k$, $J_l$ and there exists minimal common part of a subset of multi-indices present in the above sum\\ (algebraically nilpotent part)
 \item[]    or
\item[(b)] if $\xi^{J_{p}}\xi^{J_{q}}\neq 0$ for some ${J_{p}}$, ${J_{q}}$ then $\nu_{J_{p}}(t)\nu_{J_{q}}(t)$ should vanish i.e. supports of these functions should have to be disjoint, $supp\; \nu_{J_{p}}\cap supp\; \nu_{J_{q}}=\emptyset$ (functionally nilpotent part).
\end{itemize}
    Let us observe that the FNC with additional property that $\eta(t)\eta(t')=0$ $\forall t,\; t'$ has only the algebraically nilpotent part and its expansion can be written in the form
\begin{equation}\label{minfactor}
\eta(t)=\xi^{I_{min}}\left(\nu_{I_{min}}(t)+
\sum_{k, I_{k}}\nu_{I_{k+min}}(t)\xi^{I_k}\right) ,
\end{equation}
where $\xi^{I_{min}}$ is a fixed monomial in the algebra generators, common for all monomials $\xi^{I_l}$ entering the expansion (\ref{xiexpan}). From the above results we obtain the following properties of the defined by (\ref{velo}) velocities for the FON curves
\begin{equation}
\dot{\eta}^2(t)=0, \quad \eta(t)\dot{\eta}(t)=0.
\end{equation}
This is immediate conclusion from the fact that the $\dot{\eta}(t)$ inherits algebraically nilpotent part from the $\eta(t)$ and terms present in the functionally nilpotent part of the $\eta(t)$ do not contribute to the $\dot{\eta}^2(t)$ or $\eta(t)\dot{\eta}(t)$ because $supp\; \dot{\nu}(t)\subset supp\; \nu(t)$, $\nu\in C^m(\mathbb{R})$.
Let us recall that in the super-space for the odd valued curve $\theta(t)$ we always have that $\theta^2(t)=0$, $\dot{\theta}^2(t)=0$ and $\theta\dot{\theta}\neq 0$ in general. It is worth noting that we can introduce here the chain rule
\begin{equation}
\frac{d}{dt}f(\eta(t))=\dot{\eta}(t)
\partial_{\eta}f(\eta(t))
\end{equation}
which, in view of the above result, is compatible with the modified Leibniz rule (\ref{mleib}) for the $\frac{\partial}{\partial \eta}$ and the usual Leibniz rule for $\frac{d}{dt}$ when applied to $\eta^2(t)$.

%%%%%%%
To describe formalism for the many nilpotent commuting variables let us
take  $\eta_i$, $i=1,2,\dots, n$  from the set of independent
nilpotent first order  elements $\eta_i\in\tilde{\mathcal{D}}_n$
\begin{equation}
(\eta^i)^2=0 \;\; \forall \,  i, \quad \eta^1\cdot \eta^2 \cdot
\dots \eta^n\neq 0
\end{equation}
and $\vec{\eta}=(\eta^1, \eta^2, \dots \eta^n)$. The function $f(\vec{\eta})\in \mathcal{F}[\vec{\eta}]$ of
$n$ $\eta$-variables is by
\begin{equation}\label{expan}
f(\vec{\eta})=\sum_{k=0, I_k}^n f_{I_k}\eta^{I_k},
\end{equation}
where the $I_k$ denotes a strictly ordered multi-index and $f_{I_k}\in \mathcal{N}$ are constant elements. When the function $f$ depends also on the
$x\in\mathcal{K}^n$, then $f_{I_k}:\mathcal{K}^n\mapsto
\mathcal{N}$, $f\in\mathcal{F}[x,\vec{\eta}]$.

The $\eta$-derivative can be defined analogously to the superderivative
\begin{equation}
\partial_i\eta^j=\delta_i^j, \quad \partial_i 1=0,
\end{equation}
where
\begin{equation}
\partial_j=\frac{\partial}{\partial \eta^j}
\end{equation}
Moreover
\begin{equation}
\partial_i\partial_j=
\partial_j\partial_i
\end{equation}
The conventional Leibniz rule is not valid,
instead  we have the following relation
\begin{equation}\label{leib}
\partial_i(f\cdot g)= \partial_i f\cdot g + f \cdot \partial_i g -
2\eta_i\partial_i f\partial_i g
\end{equation}
One can check that such derivative has the following properties
\begin{enumerate}
\item
\begin{equation}
\partial_i(\eta_i f)= f-\eta_i\partial_i f
\end{equation}
\item
\begin{equation}
[\partial_i, \eta_i]_-=1-2\eta_i\partial_i,\quad [\partial_i,
\eta_i]_+=1
\end{equation}
\item
\begin{equation}
\nabla_i(fg)=\nabla_i fg+f\nabla_i g,\quad \mbox{for} \quad
\nabla_i=\eta_i\partial_i\quad \mbox{(no sum)}
\end{equation}
\end{enumerate}
%%%%%%%%%%%%%%%%%%%%%%%%%%%%%%%%
Next operation we shall need is the $\eta$-integration.
The $\eta$-integral is given by the following contractions \cite{amf-n2}
\begin{equation}
\int\eta_i d\eta_j=\delta_{ij}, \quad \int d\eta_i=0
\end{equation}
and has the following properties
\begin{enumerate}
\item
\begin{equation}
\int \vec{\eta}d\vec{\eta}=1, \quad
\vec{\eta}=\eta_1\eta_2\dots\eta_n,\quad
d\vec{\eta}=d\eta_1d\eta_2\dots d\eta_n
\end{equation}
\item
\begin{equation}\label{intder}
\int\partial_i f(\vec{\eta})d\eta_i=0,\quad \mbox{and}\quad
\int\partial_i f(\vec{\eta})d\vec{\eta}=0,
\end{equation}
where $f(\vec{\eta})=f(\eta_1,\eta_2,\dots,\eta_n)$
\item (the integration by part)
\begin{equation}\label{bypart}
\left(\int fd\eta_i\right)\left(\int
g\,d\eta_i\right)=\frac{1}{2}\left( \int(\partial_i f)\cdot g
d\eta_i +  \int f\cdot(\partial_i g)d\eta_i \right)
\end{equation}
\item For a matrix $A$ representing permutation and scaling
transformation, $\vec{\eta}=A\vec{\eta}'$ we have
\begin{equation}
\int f(\vec{\eta})d\vec{\eta}=(Per \,A)^{-1}\int
f(A\vec{\eta}')d\vec{\eta}',
\end{equation}
where $Per\,A$ is the permanent of the matrix $A$.
\item let $B$ be a $n\times n$ matrix then
\begin{equation}
\int e^{\eta B\eta'}d\vec{\eta}d\vec{\eta'}=Per(B)
\end{equation}
The last formula was considered already by Palumbo \cite{pal-1}.
\end{enumerate}

\section{Nilpotent mechanics}
Having above elements of the differential calculus for nilpotent commuting variables we can try to introduce the formalism of classical mechanics for nilpotent systems.
For the configuration space  for nontrivial model we will take the free $\mathcal{N}$-bimodule
$\mathbb{V}_{\mathcal{N}}$ with the $\mathcal{N}$-valued $s$-form.
\begin{equation}
s:\mathbb{V}_{\mathcal{N}}\times\mathbb{V}_{\mathcal{N}}\mapsto
\mathcal{N},
\end{equation}
where $s$-form \cite{amf-n1, amf-n2} is symmetric, non-degenerate strictly traceless f, namely
\begin{equation}
s=\left(
\begin{array}{ll}
0&\mathbb{I}_n\\
\mathbb{I}_n&0
\end{array}
\right) ;\quad s^2=\mathbb{I}_{2n}; \quad s^T=s; \quad
Tr(s)|_k=0, \, k=1,2, \dots,2n
\end{equation}
The property of being strictly traceless is dictated by the demand of having nontrivial "quadratic" expressions in $\eta^s$s. Diagonalized symmetric form would give a trivial scalar product for $\eta$-vectors. Obviously for the $s$-geometry we do not have a $GL(n)$-covariance cf. \cite{amf-n2}.\\
The Lagrangian
\begin{equation}
L=\frac{m}{2}s(\dot{\eta},\dot{\eta})-V(\eta)
\end{equation}
can be taken with the quadratic potential defining the $\eta$-oscillator
\begin{equation}
V(\eta)=\frac{m\omega}{2}s(\eta,\eta)=\frac{m\omega}{2}s_{ij}\eta^i\eta^j.
\end{equation}
Using the $\eta$-valued action of the form
\begin{equation}
I[\eta^i, \dot{\eta}^i; \alpha]=\int_{t_1}^{t_2} L(\eta^i(t,
\alpha), \dot{\eta}^i(t, \alpha))dt, \quad\quad
\alpha\in\mathbb{R}
\end{equation}
one can consider $\mathcal{N}$-valued variations
\begin{eqnarray}\nonumber
\eta^i(t, \alpha)&=&\eta^i(t)+\alpha\zeta^i(t), \quad
\zeta^i(t_1)=\zeta^i(t_2)=0\nonumber\\
\dot{\eta}^i(t, \alpha)&=&\dot{\eta}^i(t)+\alpha\dot{\zeta}^i(t)\nonumber\\
\eta^i(t)^2&=&\zeta^i(t)^2=0, \quad
\dot{\eta}^i(t)^2=\dot{\zeta}^i (t)^2=0\nonumber
\end{eqnarray}
to derive the analogs of the Euler-Lagrange equations. Because $\eta$-variations
$\eta^i(t, \alpha)^2\neq 0$ in general therefore we have two kinds of the equations of motion i.e.
\begin{eqnarray}
\mbox{EL$^{(a)}$:}\quad&&\frac{\partial
L}{\partial\eta^k}-\frac{d}{dt}(\frac{\partial
L}{\partial\dot{\eta}^k})=0,\quad
\mbox{for}\quad\eta^k(\alpha)^2\neq 0)\nonumber\\
\mbox{EL$^{(b)}$:}\quad&&\frac{\partial
L}{\partial\eta^k}-(\frac{d}{dt}-2\dot{\eta}^k\frac{\partial}{\partial\eta^k})\frac{\partial
L}{\partial\dot{\eta}^k}=0, \quad
\mbox{for}\quad\eta^k(\alpha)^2=0\nonumber
\end{eqnarray}
The modified Leibniz roule, which is used on derivation of above formulas, does not modify the shape of the equations.\\
The Hamiltonian formulation of the $\eta$-mechanics is influenced by the modified Leibniz rule. One introduces canonical momenta
\begin{equation}
p_k=\frac{\partial L}{\partial \dot{\eta}^k}
\end{equation}
and the Hamiltonian
\begin{equation}
H= \sum_k p_k\dot{\eta}^k - L.
\end{equation}
It appears \cite{amf-n2} that there are generalized Hamilton's equations
\begin{eqnarray}
      \dot{p}_k = -&\frac{\partial H}{\partial \eta^k}& \nonumber\\
    \dot{\eta^k} = &\frac{\partial H}{\partial p_k}&\nonumber
  \end{eqnarray}
However, the extension of the time derivative to the phase space
$\mathcal{P}_{\mathcal{N}}\ni f(\eta, p)$
\begin{equation}
\frac{d}{dt}=\partial_t+\sum_k\dot{\eta}^k\partial_k+\sum_k\dot{p}_k\bar{\partial}^k,\quad
\mbox{where}\,\,\,\partial_k=\frac{\partial}{\partial\eta^k}
\,\,\,\mbox{and}\,\,\,\bar{\partial}^k=\frac{\partial}{\partial
p_k}
\end{equation}
is not a usual differential operator.
Denoting
\begin{equation}
\nabla_i=\eta_i\partial_i, \quad
\bar{\nabla}^i=p^i\bar{\partial}^i, \quad \mbox{(no summation!)}
\end{equation}
Using equations of motion we get for time derivative
\begin{equation}\label{2pb}
\frac{d}{dt}f(\eta, p)=\sum_k(\bar{\partial}^k H \cdot\partial_k
f(\eta, p)-
\partial_k H\cdot\bar{\partial}^kf(\eta, p))
\end{equation}
and for product of functions
\begin{equation}
\frac{d}{dt}(g(\eta, p)\cdot h(\eta, p))=\dot{g}\cdot
h+g\cdot\dot{h}-2\sum_k(\bar{\partial}^k H \cdot\nabla_k
g\cdot\partial_k h -
\partial_k H \cdot\bar{\nabla}^k g\cdot \bar{\partial}^k h)
\end{equation}
Formula (\ref{2pb}) suggests the definition of the $\eta$-Poisson brackets:
\begin{equation}
\{f(\eta, p), g(\eta, p)\}_0=\sum_k(\bar{\partial}^k f(\eta, p)
\cdot\partial_k g(\eta, p)-
\partial_k f(\eta, p)\cdot\bar{\partial}^k g(\eta, p))
\end{equation}
These brackets have  the following properties
\begin{eqnarray}
&\mbox{(i)}&\quad\{f,g\}_0=-\{g,f\}_0\nonumber\\
&\mbox{(ii)}&\quad\{f_1+f_2,g\}_0=\{f_1,g\}_0+\{f_2,g\}_0\nonumber\\
&\mbox{(iii)}&\quad\{f,g\cdot h\}_0=\{f,g\}_0\cdot
h+g\cdot\{f,h\}_0 -2\diamondsuit(f|g,h)\nonumber\\
&\mbox{(iv)}&\quad\sum_{cycl}\{\{f,g\}, h\}_0\}_0=J(f, g, h) \nonumber
\end{eqnarray}
where the para-Leibniz term is of the form
\begin{equation}
\diamondsuit(f|g,h)=\sum_k(\bar{\partial}^k f \cdot{\nabla}_k
g\cdot\partial_k h -
\partial_k f \cdot\bar{\nabla}^k g\cdot \bar{\partial}^k h).
\end{equation}
The Jacobiator, a skew-symmetric operator $J$ is given by the formula
\begin{equation}
J(f, g, h)=2\sum_{cycl}\sum_{i}(\eta^i\{{\partial}_i f, {\partial}_i g\}\bar{\partial}^i h-p_i\{\bar{\partial}^i f, \bar{\partial}^i g\}
{\partial}_i h)
\end{equation}
It is a non-vanishing expression and spoils the Jacobi identity, moreover the Malcev identity is not satisfied as well.
Namely,
\begin{equation}
\{J(f, g, h), f\}\neq J(f, g, \{f, h\}).
\end{equation}
\section{Final remarks}
The nilpotent fuzz is very fruitfull notion and in the case of Grassmannian realization gives well known language of supermathematics intensively developed in last two decades. One can consider also a version of the nilpotent fuzz realized by means of nilpotent commuting variables. In some respects it resembles the Grassmannian case, but is essentialy different due to the lack of conventional Leibniz rule. Generalization of the classical mechanics presented here shows some new features. Such result can be viewed as natural, having in mind  a possible interpretation of the new commuting nilpotent objects. While Grassmannian variables are related to the fundamental anticommuting spinorial fields describing fermions, the new commuting nilpotent fields
are supposed to describe the composite bosonic objects (e.g. bilinear composities of fermions) like: the Cooper pairs, density fluctuations in the Tomonaga model or spin waves in ferro-antiferromagnetic model \cite{pal-1, pal-2, pal-3}. Such variables give rise to the new approach to bosonization in relativistic field theories \cite{bar-pal, car-pal}. The formalism under development allows to analyse the models of composite objects directly without referring to the
fundamental constituents. The application of the commuting nilpotent variables to the LCFT shows also their practical role in computing entities essential for the other field theories.

\end{document}